\documentclass[pra,twocolumn,amsmath,amssymb,floatfix,reprint]{revtex4-1}

\usepackage{graphicx}
\usepackage{dcolumn}
\usepackage{bm}
\usepackage[usenames,dvipsnames]{xcolor}
\usepackage{mciteplus}
\usepackage{braket}
\usepackage{ulem}
\DeclareMathOperator{\sgn}{sgn}
\begin{document}

\title{Spatio-temporal detection of Kelvin waves in 
    quantum turbulence simulations}
\author{P.~Clark di Leoni$^1$, P.D.~Mininni$^1$, \& M.E.~Brachet$^2$}
\affiliation{$^1$Departamento de F\'\i sica, Facultad de Ciencias
    Exactas y Naturales, Universidad de Buenos Aires and IFIBA, CONICET,
    Ciudad Universitaria, 1428 Buenos Aires, Argentina.\\
                  $^2$Laboratoire de Physique Statistique de l'Ecole Normale
    Sup\'erieure associ\'e au CNRS et aux Universit\'es Paris 6 et 7, 24
    Rue Lhomond, 75237 Paris Cedex 05, France.}
\date{\today}
\begin{abstract}
    We present evidence of Kelvin excitations in space-time 
    resolved spectra of numerical simulations of quantum turbulence. 
    Kelvin waves are transverse and circularly polarized waves that 
    propagate along quantized vortices, for which the restitutive 
    force is the tension of the vortex line, and which play an 
    important role in theories of superfluid turbulence. We use 
    the Gross-Pitaevskii equation to model quantum flows, letting 
    an initial array of well-organized vortices develop into a 
    turbulent bundle of intertwined vortex filaments. By achieving
    high spatial and  temporal resolution we are able to calculate 
    space-time resolved mass density and kinetic energy spectra. 
    Evidence of Kelvin and sound waves is clear in both spectra. 
    Identification of the waves allows us to extract the spatial
    spectrum of Kelvin waves, clarifying their role in the transfer of
    energy.
% This suggests that measuring the spectrum of density in experiments
% could prove useful to identify and extract Kelvin waves.
\end{abstract}
\maketitle

% Introduction 
\section{Introduction} 

Quantum turbulence is the chaotic and erratic spatio-temporal behavior
observed in superfluids and Bose-Einstein condensates (BECs) 
\cite{Vinen02,Skrbek12}. Its motion is characterized by the interaction
between vortex filaments, where all vorticity is concentrated and which
have quantized circulation \cite{Donnelly}.  As superfluids have no
viscosity, quantum turbulence has garnered much attention from the
classical turbulence community, for it could provide insight into
extremely developed turbulence.  But it has also been a matter of 
debate how similar quantum and classical turbulence actually are. 
One of the characteristical features of turbulence, the existence of a 
Kolmogorov energy spectrum, has been confirmed in 
superfluids \cite{Maurer98,Salort12}. However, the physical 
mechanisms behind this spectrum are not completely understood. 
Moreover, other features, such as velocity statistics, appear to be 
different between classical and quantum turbulence 
\cite{Paoletti08,White10}.

An interesting property of classical and quantum vortex lines, first 
predicted by Lord Kelvin \cite{Thomson80}, is that waves can
propagate along them when the line is subjected to helical deformations.
These waves are known as Kelvin waves. In quantum fluids they are
believed to play a crucial role in the turbulent energy cascade
\cite{Kozik04,Lvov10,Boue11,Boue15}, where energy is transfered from
large to small scales and it is ultimately dissipated by phonon emission
\cite{Nore97b,Vinen03}: while at large scales vortex interaction and
reconnection \cite{Meichle12} dominate the transfer of energy, at small
scales Kelvin waves are believed to interact nonlinearly exciting
fluctuations at even smaller scales. They play a role in various
problems in classical fluid dynamics \cite{Maxworthy73,Kleckner13}, but
not in classical turbulence, being the turnover time or the
period of waves particular to that system (such as inertial waves in a
rotating flow, or Alfv\'en waves in a magnetofluid) the only relevant
timescales \cite{Chen89,Clark14}. As a result, detection of Kelvin waves
in a disorganized superfluid flow is considered a signature of quantum 
turbulence \cite{Paoletti08,Fonda14}. 

There are several theoretical studies of Kelvin wave turbulence in
superfluids, whose main focus is to understand nonlinear interactions
and the resulting energy spectrum. Particular attention is given to the
latter, as predictions for the energy spectrum differ. Using wave
turbulence theory \cite{Nazarenko} different teams of researchers
arrived to different answers, one being that the energy spectrum scales
with wavenumber as $\sim k^{-7/5}$ \cite{Kozik04}, and the other that it
goes as $~\sim k^{-5/3}$ \cite{Lvov10} (it should also be noted that
    some authors claim that Kelvin waves play a neglible role in
    the spatial energy spectrum \cite{Nemirovskii12}). Recent numerical
studies yield results compatible with a $\sim k^{-5/3}$ spectrum
\cite{Baggaley11,Krstulovic12,Baggaley14,Kondaurova14}.  However,
studies of wave dynamics usually have a single or a few vortices, either
using vortex line dynamics simulations \cite{Baggaley14}, or solving the
Gross-Pitaevskii equation (GPE) \cite{Berloff14}. In the laboratory
Kelvin waves have been detected in superfluid helium using submicron ice
particles as tracers \cite{Fonda14}; the analysis focused on selected
reconnection events and the subsequent emission of Kelvin waves.
Finally, Kelvin waves and other vortex wave modes have also been studied
and identified in BECs \cite{Bretin03,Fetter04,Simula08,Simula10}.

Our aim is to study Kelvin waves in quantum turbulence.  As mentioned
above, numerical studies of quantum turbulence have two main approaches.
One is to simulate quantum vortex lines dynamics
\cite{Schwarz85,Kondaurova14}; in this approach the velocity field
outside the vortex lines is given by the Biot-Savart law and
reconnection events are performed {\it ad-hoc}. The approach we use
solves the GPE, an equation for the evolution of the wavefunction $\psi$
for a system of bosons.  A Kolmogorov spectrum $\sim k^{-5/3}$ was
obtained in simulations of the GPE in
\cite{Nore97a,Nore97b,Kobayashi05,Yepez09}.  We consider a flow that
shares similarities with the von K\'arm\'an flow generated in recent
experiments with superfluid helium \cite{Rousset14}. We extract and save
four-dimensional information to compute the spatio-temporal spectrum of
quantum turbulence, where the presence of sound and Kelvin waves is
evident. Previous studies using a similar technique observed sound waves
\cite{Nazarenko06,Proment09}, and Kelvin waves in a single and straight
vortex filament \cite{Baggaley14}. Our study is done in a
three-dimensional highly turbulent environment, with a large number of
vortices.

% Equations and numerical setup 
\section{The Gross-Pitaevskii equation}

The GPE describes the evolution of the wavefunction $\psi$ of a field 
of weakly interacting bosons of mass $m$,
\begin{align}
    \label{gpe}
    i \hbar \frac{\partial \psi}{\partial t} = -
    \frac{\hbar^2}{2m}\nabla^2 \psi + g \vert \psi \vert^2 \psi ,
\end{align}
where $g$ is proportional to the scattering length. By means of the
Madelung transformation
\begin{align}
    \label{madelung}
    \psi ({\bf r},t) = \sqrt{\frac{\rho({\bf r},t)}{m}} e^{i
        m \phi ({\bf r},t)/\hbar} ,
\end{align}
where $\rho$ is the density of particles and the phase $\phi$ can be
associated with a velocity by ${\bf v}={\bm \nabla} \phi$, one obtains
a hydrodynamic description of the system \cite{Proukakis08},
\begin{gather}
    \label{mass}
    \frac{\partial \rho}{\partial t} + {\bm \nabla} \cdot (\rho {\bf v})
        = 0,
    \\
    \label{vel}
    \frac{\partial {\bf v}}{\partial t} + {\bf v}\cdot {\bm \nabla} {\bf
        v} = - \frac{g}{m^2} {\bm \nabla} \rho + \frac{\hbar^2}{2 m^2}
    {\bm \nabla} \left( \frac{\nabla^2 \sqrt{\rho}}{\sqrt{\rho}}
    \right).
\end{gather}
These equations are similar to the Euler equations for a classical and
compressible barotropic fluid, except for the extra second term on the
r.h.s.~of Eq.~\eqref{vel} which is referred to as the ``quantum 
pressure''. By solving the GPE we get the full three dimensional
velocity and density fields, and compared to other methods
\cite{Nemirovskii13}, we do not need to reconstruct them from the vortex
tangle configuration.

The hydrodynamic description allows us to define a quantity akin to the
classical kinetic energy of a fluid, namely $E_k = \rho v^2/2$.  Note
that this is only a fraction of the total energy density, given by
$\hbar^2 \vert \nabla \psi \vert^2 /2m + g \vert \psi \vert^4 /2$. For
simplicity, we will refer to the classical kinetic energy just as the
kinetic energy. It can be further decomposed into an incompressible
component $E^i_k$ and a compressible component $E^c_k$, by decomposing
the velocity field into irrotational and solenoidal components. The two
components will be useful to discriminate between sound waves and other
excitations in the fluid. More details on the energy decompositions are
given
below (see also \cite{Nore97a,Krstulovic11}).

This system can have sound waves which 
follow the Bogoliubov dispersion relation \cite{Proukakis08},
\begin{align}
    \label{bogoliubov}
    \omega_B (k) = k \sqrt{ c^2 + \frac{c^2 \xi^2}{2} k^2} ,
\end{align}
where the speed of sound is $c=\sqrt{g|\psi|^2/m}$ and the coherence 
length is $\xi=\sqrt{\hbar^2/(2m|\psi|^2g)}$
\cite{Proukakis08,Nore97a}, and Kelvin waves \cite{Donnelly} which 
follow the dispersion relation
\begin{align}
    \label{kelvin}
    \omega_K (k) = \frac{2 c \xi}{\sqrt{2} a^2} \left(1 \pm
        \sqrt{1 + k a \frac{K_0(ka)}{K_1(ka)}}
    \right),
\end{align}
where $a$ is the vortex core radius, and $K_0$ and $K_1$ are
modified Bessel functions. The dispersion relation is quadratic
in the small $k$ limit, while for large $k$ it is linear.

To solve numerically Eq.~\eqref{gpe} we use GHOST \cite{Mininni11}, a
highly parallel code which uses a pseudospectral method to compute
spatial derivatives, fourth-order Runge-Kutta to compute time
derivatives, and can solve PDEs in Cartesian periodic grids. We use
$512^3$ grid points and the ``2/3 rule'' for de-aliasing. The speed of
sound is $c=2$ and the coherence length is $\xi = 0.1/(8\sqrt{2})$ in
dimensionless units in a three-dimensional box of length $L=2\pi$.
Quantities are made dimensionless using characteristic length $L_0$,
velocity $U_0$, and mean density $\rho_0$ (see \cite{Nore97a}). These 
parameters result in an intervortex distance $\ell$ such that 
$k_\ell = 2\pi/ \ell \approx 10$ (see \cite{Nore97a,Nore97b}), and in 
a vortex core radius $a \approx 2 \xi$ as measured directly from the 
full width at half maximum of the mass density profile. Also, 
the quantum of circulation $h/m$ in dimensionless units is 
given by $4\pi c \xi/\sqrt{2}=0.05\pi$.

As initial condition we use the Taylor-Green flow, which results in a 
set of vortex loops in two counter-rotating large scale eddies with 
turnover time of order unity, and whose geometry mimics the von K\'arm\'an 
flow \cite{Nore97a}. The von K\'arm\'an flow is used in recent experiments 
with two counter-rotating propellers such as SHREK (the {\it Superfluid
  High REynolds von K\'arm\'an experiment}) \cite{Rousset14}, and has 
been used in the past to measure Kolmogorov spectra in superfluids 
\cite{Maurer98}. For more information about the generation of
the initial conditions see Appendix~\ref{app_a}. We let the simulation run for 
$\approx 20$ large-scale turnover times, so as to get good statistics 
on the slowest waves in the system. To also resolve the fastest waves, 
we use a very high output cadence, storing one output of the 
wavefunction every half period of the fastest waves (i.e., the sound 
waves for the maximum wavenumber in the spatial domain). This 
very high temporal resolution allows us to properly calculate space 
and time resolved spectra.

% Characterization of the flow 
\section{Characterization of the flow} 

\begin{figure}
    \centering
    \includegraphics[width=8.5cm]{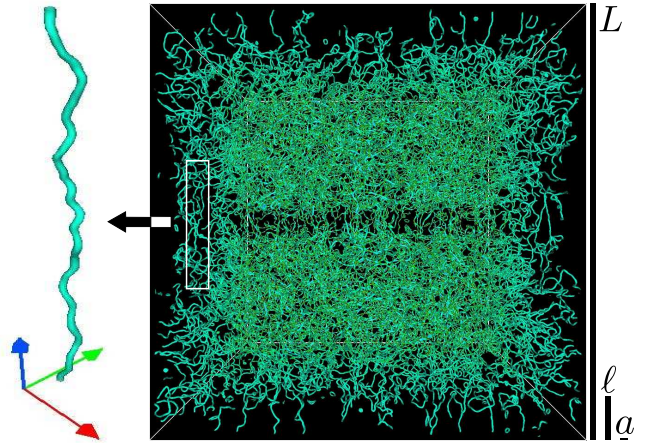}
    \caption{({\it Color online}) Three-dimensional rendering of the
        mass density. The lines, corresponding to regions of low
        density, are associated with quantized vortices. Typical length
        scales are indicated by the black bars (see text for
        description). On the left we show a zoom into a single
            vortex; helical perturbations propagating along
            it can be identified. Due to the highly
        turbulent nature of the flow, identifying waves here is like
        looking for needles in a haystack.}
    \label{vapor}
\end{figure}

Figure \ref{vapor} shows a three-dimensional rendering of the density
field $\rho$. The visualization is done using the software VAPOR
\cite{Clyne07}, and only regions with low density (indicating
topological defects associated with quantized vortices) are shown. The
field evolves from a well ordered structure to a bundle of intertwined
and structurally rich vortices.  This evolution is associated with the
development of reconnection events, which become prominent at
$t\approx5$ and start to subside after $t\approx 10$. The analysis below
is done during this period, as afterwards too much of the total energy
has decayed into phonons.  Tracking reconnection events and the
subsequent generation of Kelvin waves in this flow is not an easy task.
However, the identification of helical waves from spatial observations
is possible if single vortices or single reconnection events are
followed in time (see, e.g., \cite{Fonda14}).

In Fig.~\ref{energy_evolA} we present the evolution of the different
components of the total energy in the simulation. As mentioned above, 
the classical kinetic energy density is defined as
\begin{align}
    E_k = \frac{1}{2} \rho v^2,
\end{align}
The remaining components of the total energy density in the system are
the so-called quantum energy
\begin{align}
    E_q = \frac{\hbar^2}{2 m^2} (\nabla \sqrt{\rho})^2,
\end{align}
and the potential (or internal) energy
\begin{align}
    E_p =  \frac{g}{2 m^2} \rho^2 .
\end{align}
A detailed analysis and derivation of each component of the total
energy density can be found, e.g., in \cite{Krstulovic11}.

From Fig.~\ref{energy_evolA} different regimes can be identified in
the evolution. In the first stage, up to $t\approx5$, the
incompressible kinetic energy oscillates around a mean value. 
Afterwards it decays and the other components of the energy 
grow, as reconnection of vortex lines takes place and the flow 
becomes more complex (see Fig.~1). 
After $t\approx10$ the number of reconnection events in the flow 
subsides and the growth of the compressible, potential, and 
quantum energy becomes slower. During this process the total 
energy is conserved up to the sixth significant digit. Mass and 
momentum are also conserved. Snapshots of the velocity field 
and of the density between $t\approx5$ and $\approx10$, when 
the turbulent spectrum is more developed, are thus used for the 
spatio-temporal analysis.

% Spatiotemporal analysis: finding waves amongst the turbulence 
\section{Looking for needles in a haystack}
\label{haystack}

\begin{figure}
    \centering
    \includegraphics[width=9.5cm]{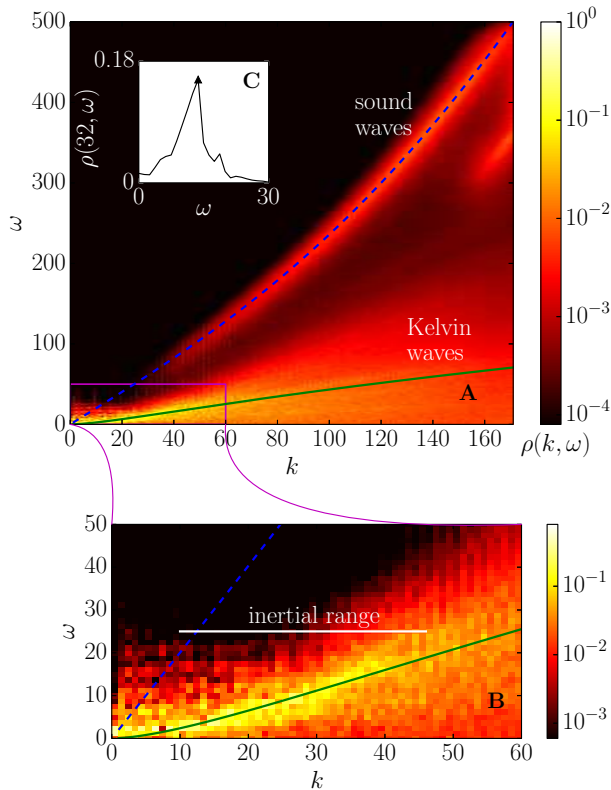}
    % \vspace{-1cm}
    \caption{({\it Color online}) Space-time resolved mass 
        spectrum, $\rho(k,\omega)$ normalized by $\rho(k)$. 
        A: The whole spectrum; the dashed (blue) line indicates  
        the sound wave dispersion relation $\omega_B(k)$, the 
        solid (green) line indicates the Kelvin wave dispersion
        relation $\omega_K(k)$. B: Close-up for small wavenumbers; 
        note the accumulation of power near the Kelvin wave modes.
        The range of wavenumbers with strong Kelvin excitations 
        are marked as ``inertial range''.  
        C: A cut of $\rho(k,\omega)$ (also normalized) for $k=32$, 
        the marker indicates the Kelvin wave frequency. At low 
        wavenumbers Kelvin waves are dominant, whereas sound 
        waves become prominent after $k\approx50$. At very 
        high wavenumbers the quantum pressure inside the vortex 
        core also leaves a trace in the spectrum.}
    \label{rhokw}
\end{figure}

 Instantaneous flow
visualization is insufficient to identify and extract all the waves in a
turbulent flow. In particular, in order to quantify their relevance in
the energy cascade, it is necessary to quantify their amplitudes as a
function of frequency and wavenumber, i.e., to calculate space-time
resolved spectra. As Kelvin waves are oscillations of the lines with
$\rho=0$, we should be able to identify their imprint in the mass
spectrum, and we therefore study it first. Figure \ref{rhokw} shows the
mass spectrum $\rho(k,\omega)$, along with a close-up for low
wavenumbers, as well as a cut of $\rho(k,\omega)$ at $k=32$. The dashed
line in Fig.~\ref{rhokw} corresponds to the Bogoliubov linear
dispersion relation of sound waves [Eq.~\eqref{bogoliubov}], while the
solid line corresponds to that of Kelvin waves [Eq.~\eqref{kelvin}].
Substantial power is concentrated along these curves. At low wavenumbers
Kelvin waves are dominant, whereas sound waves become prominent as
Kelvin waves begin to fade for $k\gtrsim50$. At very high 
wavenumbers another accumulation of energy can be seen for 
$\omega\approx 350$, which is probably due to quantum pressure 
effects inside the vortex core. Indeed, the spectrum of the quantum 
pressure also shows a bump at these wavenumbers (not shown), and 
the frequencies are compatible with those predicted for axisymmetric 
oscillations of the vortex core \cite{roberts03,Bradley12}. The
presence of Kelvin waves at low wavenumbers is the most striking
feature of the flow; these modes are probably excited by deformation 
of vortex lines by the large-scale flow, and are hard to identify by
simple inspection of the evolution of individual vortex lines. In
Appendix~\ref{app_b}, we present a benchmark sutdy of the spatio-temporal
spectrum of mass in a flow where only linear Kelvin waves were excited,
to verify the modes indentified in Fig.~\ref{rhokw} correspond to these
waves.

\begin{figure}
    \centering
    \includegraphics[width=8.2cm]{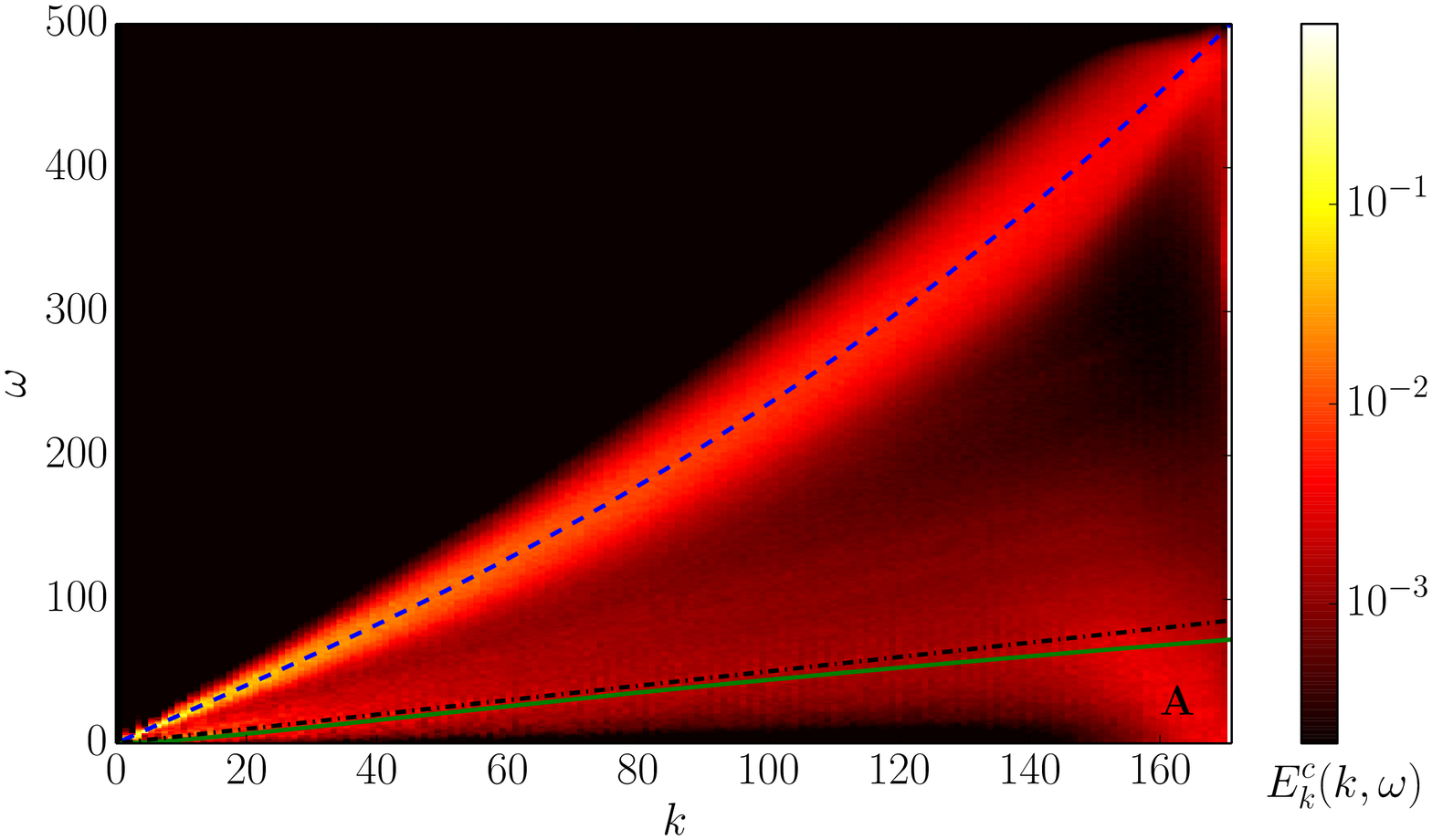}
    \\
    \includegraphics[width=8.2cm]{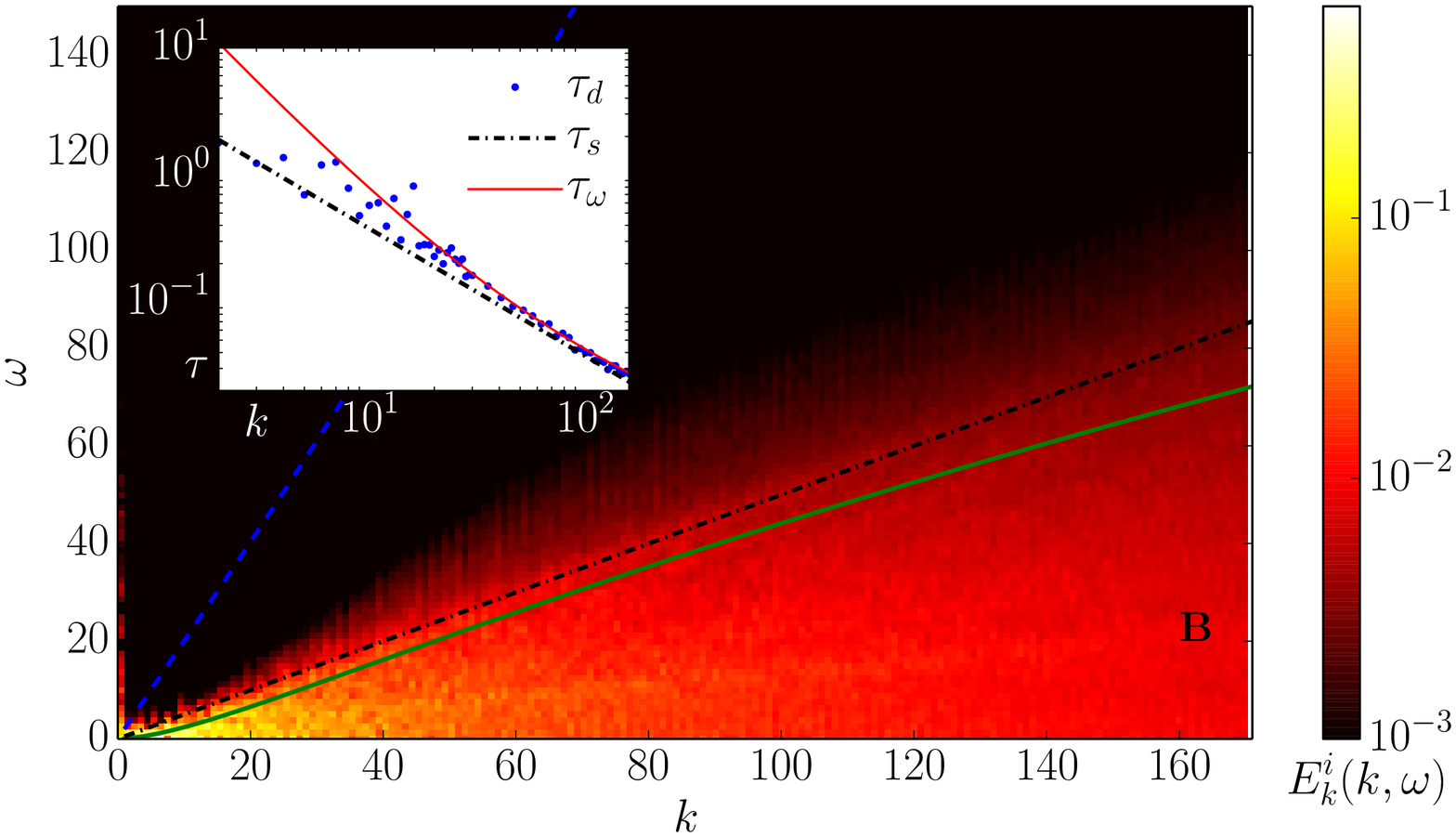}
    \caption{({\it Color online}) A: Space-time resolved compressible
        kinetic energy spectrum, $E^c_k(k,\omega)$, normalized by
        $E^c_k(k)$. B: Same for the incompressible kinetic energy
        spectrum $E^i_k(k,\omega)$. The dashed (blue) line corresponds
        to $\omega_B(k)$, and the solid (green) line corresponds to
        $\omega_K(k)$. The dash-dotted (black) line corresponds to
        sweeping, which excites all modes with frequency equal or
        smaller than $\omega = U_{rms}k$. $E^c_k(k,\omega)$ is 
        dominated by excitations around the sound wave dispersion 
        relation while $E^i_k(k,\omega)$ shows sweeping and Kelvin
        wave excitations. Inset: decorrelation time $\tau_d(k)$ 
        of individual Fourier modes, compared with the sweeping time 
        $\tau_s$ (dash-dotted black line) and the Kelvin wave period 
        $\tau_\omega$ (solid red line).}
    \label{kckw}
\end{figure}

To independently verify the presence of sound and Kelvin waves, and to
separate the multiple branches of the dispersion relation, we now consider
the spectrum of the compressible kinetic energy $E^c_k (k,\omega)$, and
of the incompressible kinetic energy $E^i_k (k,\omega)$ (see
Fig.~\ref{kckw}).  A strong accumulation of energy around modes
satisfying the relation $\omega=\omega_B(k)$ is evident in the
compressible spectrum (Fig.~\ref{kckw}A). In the incompressible
spectrum (Fig.~\ref{kckw}B) these excitations are negligible, but two
new features are found. First, strong excitations can be observed for
all wavenumbers at low frequencies. These excitations are compatible
with sweeping of the vortex cores by the large-scale flow, i.e., the
advection of small scale structures by the flow with a slow timescale
associated with the turnover time. This is an important effect in
classical turbulence where it is responsible for the temporal
decorrelation of modes \cite{Chen89,Clark14}. It results in the
smearing of the energy for all frequencies $\omega = U_{rms}k$ and
smaller (with r.m.s.~velocity $U_{rms} \approx 0.5$). Superposed to
these excitations, modes compatible with Kelvin waves can still be
observed for small wavenumbers. To further verify
    this, we calculated the decorrelation time $\tau_d$ of
    individual Fourier modes and compared it to the sweeping time 
    $\tau_s \sim 1/(U_{rms}k)$ and to the wave period $\tau_\omega \sim
    1/\omega_K(k)$ (Fig.~\ref{kckw}). The decorrelation time
    fluctuates between $\tau_s$ and $\tau_\omega$ for low wavenumbers 
    and converges towards $\tau_s$ for large wavenumbers; no other 
    relevant time scales can be observed. Other collective vortex
    motions (e.g., Tkachenko waves
    \cite{Baym03,Coddington03,Mizushima04}) may also be present 
    but masked by these two dominant timescales. A
weak trace of these effects is also present in $E^c_k(k,\omega)$. 
Kelvin waves do not contribute significant energy to the spectrum 
of classical turbulence, making the presence of energy in these modes 
a signature of quantum turbulence.

\begin{figure}
    \centering
    \includegraphics[width=8.5cm]{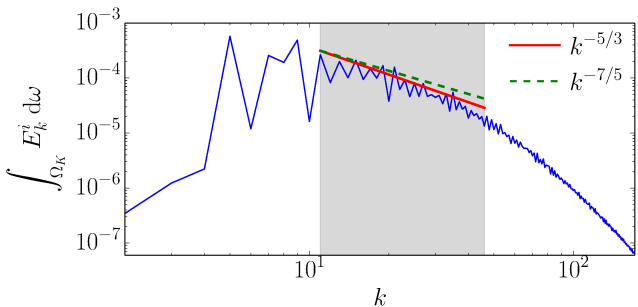}
    \vspace{-0.5cm}
    \caption{({\it Color online}) Spectrum of the incompressible
        kinetic energy associated with Kelvin wave modes, 
        $\int_{\Omega_K} E^i_k (k,\omega) \; \mathrm{d}\omega$, where 
        $\Omega_K$ are the modes neighboring $\omega_K(k)$. The
            two predictions for the Kelvin wave spectrum are shown as
            references. The shaded area corresponds to the region with
            strong Kelvin excitations identified as ``inertial range''
            in Fig.~\ref{rhokw}.B}
    \label{reduced}
\end{figure}

The role of Kelvin waves in the dynamics of quantum turbulence is
controversial. While at large scales the interaction of quantized
vortices with the flow results in advection and reconnection of vortex
lines, at scales comparable to the intervortex distance $\ell$ Kelvin
waves are believed to interact nonlinearly, exciting smaller
fluctuations that eventually loose their energy to phonons. This
transfer of energy by nonlinear interaction of wave modes can be
described by wave turbulence theories, but current predictions differ on
the shape of the energy spectrum \cite{Kozik04,Lvov10}. The
quantification of the amplitude of all Kelvin waves modes shown in
Figs.~\ref{rhokw} and Fig.~\ref{kckw} can be used to shed some light on
this problem.  We thus extract the modes centered around the dispersion
relation given by Eq.~\eqref{kelvin} with a width of $2 \sigma$ from the
wave frequency, where the dispersion $\sigma$ is estimated by fitting
the peaks in $\rho(k_i,\omega)$ such as the one shown in
Fig.~\ref{rhokw}C with a Gaussian. This subset of modes, which we call
$\Omega_K$, can be used to compute the spectrum of the incompressible
kinetic energy associated with Kelvin wave modes, $\int_{\Omega_K}
E^i_k(k,\omega) \;\mathrm{d} \omega$ (Fig.~\ref{reduced}). We verified
that the shape of the spectrum is not very sensitive to the width in
units of $\sigma$ used to define the subset $\Omega_K$, as long as it is
not too large so as to include, e.g., modes associated with sound waves.
In this spectrum, the candidate for an inertial range is observed for
$k\gtrsim k_\ell$, and its width is in agreement with the wavenumbers
for which we observe Kelvin-like excitations in Fig.~\ref{rhokw}. As the
scale separation is limited, we do not attempt to fit the data in
Fig.~\ref{reduced} and only show the two theoretical predictions for the
scaling as references.

% Conclusions 
\section{Conclusions} 

Kelvin wave turbulence is an inherently quantum
regime of superfluids.  Proper quantification of Kelvin waves is
important to understand the differences between classical and quantum
turbulence.  We presented direct evidence of the presence of Kelvin
waves in numerical simulations of quantum turbulence using the GPE. By
looking at the space-time resolved mass density spectrum, we showed
that Kelvin waves play the dominant role at scales comparable to the
intervortex distance, while sound waves are excited at smaller
scales.  Furthermore, the kinetic energy spectrum confirms the
presence of both waves in the flow. The presence of these waves in 
the spectrum can be considered a quantum signature that distinguishes 
this flow from classical turbulence.

Modes satisfying the wave dispersion relations excite a continuous
region of the spectra, with broadening around the theoretical dispersion
curves. This suggest that waves interact nonlinearly, as broadening of
the dispersion relation in turbulent flows is often the result of
nonlinear coupling and energy transfer.  However, the modes excited are
still close to the linear dispersion relations, an unexpected result as
excitations in the turbulent regime do not necessarily have small
amplitudes.  Finally, extraction of the spectrum of the incompressible
kinetic energy of modes along the dispersion relation of Kelvin waves
results in a spectrum that is compatible with predictions of weak
turbulence theories for superfluids.

\appendix
% Appendix A
\section{Preparation of the initial conditions}
\label{app_a}

The Taylor-Green initial conditions are generated by preparing a
wavefunction $\psi$ whose associated velocity field is a Taylor-Green
flow \cite{Nore97a}, given by
\begin{align*}
    v_x (x,y,z) &= \sin(x)\cos(y)\cos(z) ,
    \\
    v_y (x,y,z) &= - \cos(x)\sin(y)\cos(z) ,
    \\
    v_z (x,y,z) &= 0 .
\end{align*}
The Taylor-Green flow contained in a periodic box has properties 
that mimic the von K\'arm\'an flow driven by two counter-rotating 
impellers, and has been extensively used in simulations to 
compare with the experimental flow. The von K\'arm\'an flow is 
commonly used in the laboratory to study turbulence, including 
dynamo experiments with conducting flows, experiments to study 
Lagrangian particles, and superfluid turbulence \cite{Rousset14}.

The process to generate the initial wavefunction is described in 
great detail in \cite{Nore97a}. Here we give a brief, but nonetheless 
complete, presentation. The Taylor-Green flow can be described 
by the Clebsch potentials
\begin{align*}
    \lambda (x,y,z) &= \cos(x) \sqrt{2 \vert \cos(z) \vert} ,
    \\
    \mu (x,y,z) &= \cos(y) \sqrt{2 \vert \cos(z) \vert} \sgn[\cos(z)] ,
\end{align*}
which verify ${\bm \nabla} \times {\bf v} = {\bm \nabla} \lambda
\times {\bm \nabla} \mu$. Our use for them is that they can map a 
point in the $(\lambda,\mu)$ plane to a line in three-dimensional real
space. Now, instead of having to assemble directly a three-dimensional
wavefunction whose nodal lines match vortex lines of the velocity
field ${\bf v}$, we can pick instead a complex field $\phi
(\lambda,\mu)$ which has a zero (a defect) at the point 
$(\lambda_d,\mu_d)$. Then, the three-dimensional wavefunction
\begin{align}
    \varphi(x,y,z) = \phi (\lambda(x,y,z),\mu(x,y,z)),   
\end{align}
will be equal to zero along the line(s) defined by
$\lambda(x,y,z)=\lambda_d$ and $\mu(x,y,z)=\mu_d$. This ensures 
that the defects of $\varphi$ match vortex lines of ${\bf v}$, as
desired.

Our choice for $\phi$ is
\begin{align*}
    \phi(\lambda,\mu) = &\phi_e\left(\lambda-\frac{1}{\sqrt{2}},\mu\right)
                         \phi_e\left(\lambda,\mu-\frac{1}{\sqrt{2}}\right)
                        \\
                         &\times
                         \phi_e\left(\lambda+\frac{1}{\sqrt{2}},\mu\right)
                         \phi_e\left(\lambda,\mu+\frac{1}{\sqrt{2}}\right),
\end{align*}
with
\begin{align*}
    \phi_e (\lambda,\mu) =\frac{(\lambda+i\mu)}{\sqrt{\lambda^2+\mu^2}}
    {\tanh \left( \frac{\sqrt{\lambda^2+\mu^2}}{\sqrt{2}\xi} \right)} .
\end{align*}
As $\phi_e$ has one simple zero at the origin, $\phi$ will have four
simple zeros in the region $[0,\pi]\times[0,\pi]$, resulting 
in four nodal lines in three-dimensions. But as the circulation of 
the velocity ${\bf v}$ must match the circulation generated by the 
vortex lines (whose circulation is in turn quantized and fixed by the 
parameters of the simulation), we must change the multiplicity of 
the nodal lines to match both values. Using that the total 
circulation of ${\bf v}$ is $\Gamma=8$, that the quantum of
circulation is $4\pi \alpha$ (with $\alpha = \hbar /2m$), and that 
we want multiples of four nodal lines, we write
\begin{align}
    \label{ci_argle}
    \psi(x,y,z) = \varphi (x,y,z)^{\{1/(2\pi\alpha)\}} ,
\end{align}
where the brackets $\{\}$ denote the integer part.

This wavefunction is then evolved in time for over $30$ turnover
times under the dynamics of the Advective Real Ginzburg-Landau 
Equation (ARGLE),
\begin{align}
    \label{argle}
     \frac{\partial \psi}{\partial t} =\frac{c \xi}{\sqrt{2}} \nabla^2
     \psi + \frac{c}{\sqrt{2} \xi} \left( \vert \psi \vert^2 \psi -
     \frac{\vert \psi \vert^4}{2} \right) \nonumber \\ 
     -i {\bf v} \cdot \nabla \psi
     -\frac{\sqrt{2}{\bf v}^2}{4 c \xi} \psi .
\end{align}
This allows us to reduce the acoustic energy present in $\psi$, 
as solutions of the ARGLE converge to states which 
are solutions of the GPE with minimal energy in acoustic modes. Also,
evolving the system under the ARGLE improves the matching of the
circulations. The resulting wavefunction is finally used as the initial
condition for the GPE.

\begin{figure}
    \centering
    \includegraphics[width=8.5cm]{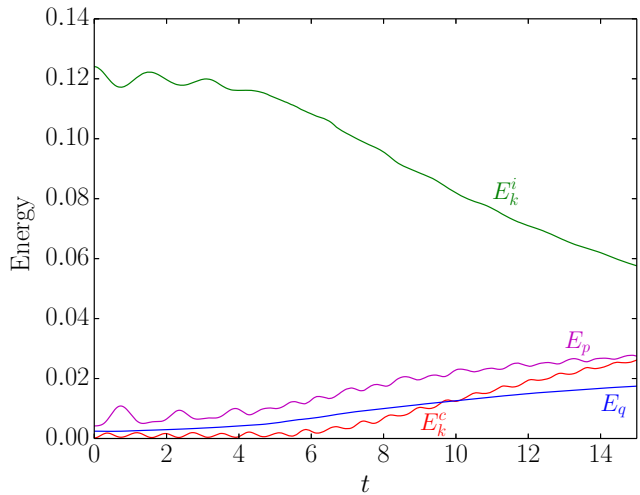}
    \caption{Time evolution of the incompressible kinetic energy 
      $E^i_k$, of the compressible kinetic energy $E^c_k$, of the
      potential energy $E_p$, and of the quantum energy $E_q$.}
    \label{energy_evolA}
\end{figure}

\section{Analysis for helical perturbations}
\label{app_b}

\begin{figure}[h]
   \centering
   \includegraphics[width=8.5cm]{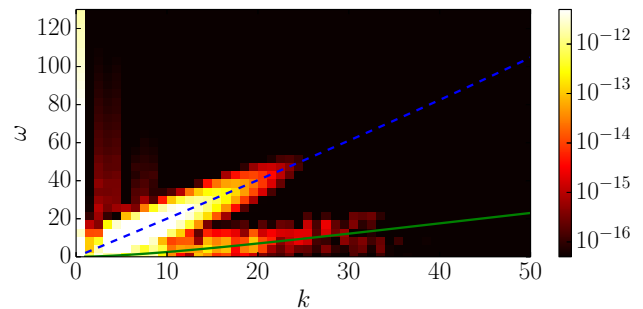}
   \caption{Space-time resolved mass density spectrum 
        $\rho(k,\omega)$ for a simulation with four straight vortices 
        perturbed with small helical displacements between $k=3$ and 
        $k=30$. The dashed (blue) line indicates the sound wave 
        dispersion relation $\omega_B(k)$, and the solid (green) line 
        indicates the Kelvin wave dispersion relation $\omega_K(k)$.}
   \label{ek}
\end{figure}

As a reference, and for comparison with the spectra presented 
in Sec.~\ref{haystack} for the Taylor-Green initial conditions, we present
here the space-time resolved spectrum for a system of only four 
straight vortices at rest, perturbed with small helical perturbations 
so as to excite linear Kelvin waves. The spatial resolution and
parameters of the simulation were the same as in the simulation in the 
main text. Initial conditions were prepared as in \cite{Krstulovic12}, and
the vortices were perturbed with a superposition of small helical 
displacements between wavenumbers $k=3$ and $k=30$ (see also 
\cite{Krstulovic12}). The resulting space-time resolved mass density
spectrum is shown in Fig.~\ref{ek}. The perturbation excites strong 
sound waves, but modes compatible with the dispersion relation of 
Kelvin waves can be identified in the range of wavenumbers excited 
by the small initial perturbation. Moreover, the dispersion relation
of these modes is compatible with the one observed in the main text 
for the fully developed turbulent flow.

\begin{acknowledgments}
The authors acknowledge support from Grant No. ECOS-Sud A13E01. 
PCdL and PDM acknowledge support from Grant Nos. PIP 
11220090100825, UBACYT 20020110200359, PICT 2011-1529 and 
PICT 2011-1626. PDM acknowledges fruitful discussions with C. Rorai, 
E. Calzetta, and V. Bekeris.
\end{acknowledgments}

\bibliography{ms}

\end{document}